\documentclass[aps,prl,preprint,tightenlines,superscriptaddress,showpacs,byrevtex]{revtex4}

\usepackage{graphicx}
\usepackage{epsfig}

\newcommand{\bdk}{B^\pm\rightarrow D K^\pm}
\newcommand{\bdstk}{B^\pm\rightarrow D^{*} K^\pm}

\begin{document}

\vspace*{-3\baselineskip}
\resizebox{!}{3cm}{\includegraphics{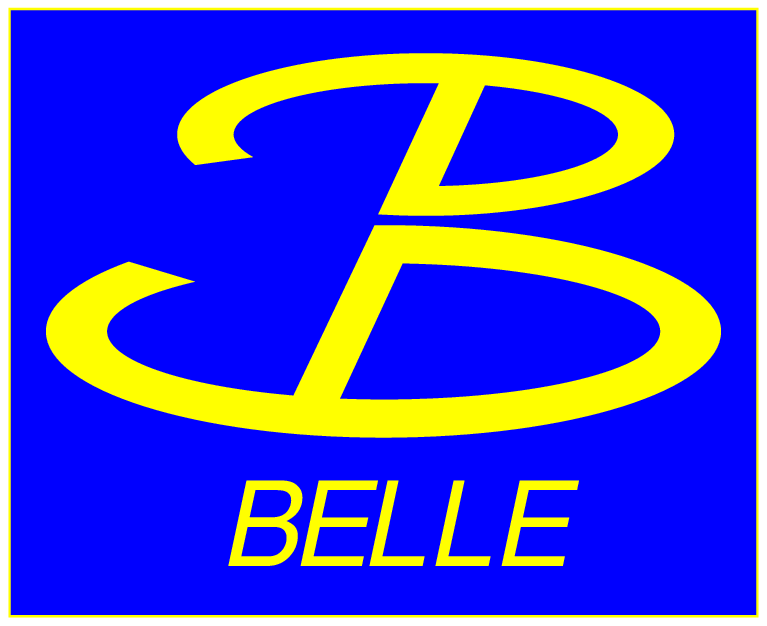}}

\preprint{\vbox{ \hbox{   }
                 \hbox{BELLE Preprint 2006-1}
                 \hbox{KEK Preprint 2005-91}
}}

\vspace*{12mm}

\title{ \quad\\[0.5cm]
{\boldmath Study of $B^\pm \rightarrow D_{CP}K^\pm$ and 
$D^{*}_{CP} K^\pm$ decays }}

%
%%% Paper:    B -> D_CP(*) K
%%% Journal:  PRD(RC)
%%% Contacts: N. Kent (kent@phys.hawaii.edu)
%%%           K. Trabelsi (karim@phys.hawaii.edu)
%%% Non-responding authors or those who said NO are commented out.
%%% ====================================================================
%%% Click the RELOAD button on your web browser to see the updated file.
%%% ====================================================================
%%% Use \input{author} to insert this material into your latex file.
%%%%% Force institutions to appear in alphabetical order when typeset.
\affiliation{Aomori University, Aomori}
\affiliation{Budker Institute of Nuclear Physics, Novosibirsk}
\affiliation{Chiba University, Chiba}
\affiliation{Chonnam National University, Kwangju}
\affiliation{University of Cincinnati, Cincinnati, Ohio 45221}
\affiliation{University of Frankfurt, Frankfurt}
\affiliation{Gyeongsang National University, Chinju}
\affiliation{University of Hawaii, Honolulu, Hawaii 96822}
\affiliation{High Energy Accelerator Research Organization (KEK), Tsukuba}
\affiliation{Hiroshima Institute of Technology, Hiroshima}
\affiliation{Institute of High Energy Physics, Chinese Academy of Sciences, Beijing}
\affiliation{Institute of High Energy Physics, Protvino}
\affiliation{Institute of High Energy Physics, Vienna}
\affiliation{Institute for Theoretical and Experimental Physics, Moscow}
\affiliation{J. Stefan Institute, Ljubljana}
\affiliation{Kanagawa University, Yokohama}
\affiliation{Korea University, Seoul}
\affiliation{Kyoto University, Kyoto}
\affiliation{Kyungpook National University, Taegu}
\affiliation{Swiss Federal Institute of Technology of Lausanne, EPFL, Lausanne}
\affiliation{University of Ljubljana, Ljubljana}
\affiliation{University of Maribor, Maribor}
\affiliation{University of Melbourne, Victoria}
\affiliation{Nagoya University, Nagoya}
\affiliation{Nara Women's University, Nara}
\affiliation{National Central University, Chung-li}
\affiliation{National United University, Miao Li}
\affiliation{Department of Physics, National Taiwan University, Taipei}
\affiliation{H. Niewodniczanski Institute of Nuclear Physics, Krakow}
\affiliation{Nippon Dental University, Niigata}
\affiliation{Niigata University, Niigata}
\affiliation{Nova Gorica Polytechnic, Nova Gorica}
\affiliation{Osaka City University, Osaka}
\affiliation{Osaka University, Osaka}
\affiliation{Panjab University, Chandigarh}
\affiliation{Peking University, Beijing}
\affiliation{University of Pittsburgh, Pittsburgh, Pennsylvania 15260}
\affiliation{Princeton University, Princeton, New Jersey 08544}
\affiliation{RIKEN BNL Research Center, Upton, New York 11973}
\affiliation{Saga University, Saga}
\affiliation{University of Science and Technology of China, Hefei}
\affiliation{Seoul National University, Seoul}
\affiliation{Shinshu University, Nagano}
\affiliation{Sungkyunkwan University, Suwon}
\affiliation{University of Sydney, Sydney NSW}
\affiliation{Tata Institute of Fundamental Research, Bombay}
\affiliation{Toho University, Funabashi}
\affiliation{Tohoku Gakuin University, Tagajo}
\affiliation{Tohoku University, Sendai}
\affiliation{Department of Physics, University of Tokyo, Tokyo}
\affiliation{Tokyo Institute of Technology, Tokyo}
\affiliation{Tokyo Metropolitan University, Tokyo}
\affiliation{Tokyo University of Agriculture and Technology, Tokyo}
\affiliation{Toyama National College of Maritime Technology, Toyama}
\affiliation{University of Tsukuba, Tsukuba}
\affiliation{Virginia Polytechnic Institute and State University, Blacksburg, Virginia 24061}
\affiliation{Yonsei University, Seoul}
  \author{K.~Abe}\affiliation{High Energy Accelerator Research Organization (KEK), Tsukuba} % KEK
  \author{K.~Abe}\affiliation{Tohoku Gakuin University, Tagajo} % TohokuGakuin
% \author{N.~Abe}\affiliation{Tokyo Institute of Technology, Tokyo} % TIT
  \author{I.~Adachi}\affiliation{High Energy Accelerator Research Organization (KEK), Tsukuba} % KEK
  \author{H.~Aihara}\affiliation{Department of Physics, University of Tokyo, Tokyo} % Tokyo
% \author{K.~Aoki}\affiliation{Nagoya University, Nagoya} % Nagoya
% \author{K.~Arinstein}\affiliation{Budker Institute of Nuclear Physics, Novosibirsk} % BINP
  \author{Y.~Asano}\affiliation{University of Tsukuba, Tsukuba} % Tsukuba
% \author{T.~Aso}\affiliation{Toyama National College of Maritime Technology, Toyama} % Toyama
% \author{V.~Aulchenko}\affiliation{Budker Institute of Nuclear Physics, Novosibirsk} % BINP
  \author{T.~Aushev}\affiliation{Institute for Theoretical and Experimental Physics, Moscow} % ITEP
  \author{T.~Aziz}\affiliation{Tata Institute of Fundamental Research, Bombay} % Tata
  \author{S.~Bahinipati}\affiliation{University of Cincinnati, Cincinnati, Ohio 45221} % Cincinnati
  \author{A.~M.~Bakich}\affiliation{University of Sydney, Sydney NSW} % Sydney
% \author{V.~Balagura}\affiliation{Institute for Theoretical and Experimental Physics, Moscow} % ITEP
% \author{Y.~Ban}\affiliation{Peking University, Beijing} % Peking
% \author{S.~Banerjee}\affiliation{Tata Institute of Fundamental Research, Bombay} % Tata
% \author{E.~Barberio}\affiliation{University of Melbourne, Victoria} % Melbourne
  \author{M.~Barbero}\affiliation{University of Hawaii, Honolulu, Hawaii 96822} % Hawaii
% \author{A.~Bay}\affiliation{Swiss Federal Institute of Technology of Lausanne, EPFL, Lausanne} % Lausanne
% ...
% Lausanne
\collaboration{The Belle Collaboration}

\begin{abstract}
\noindent We report a study of the modes $\bdk$ and $\bdstk$ where
$D^{(*)}$ decays to $CP$ eigenstates. The data sample used contains 
275 $\times 10^6$ $B\bar{B}$ events at the $\Upsilon(4S)$
resonance collected by the Belle detector at the KEKB energy-asymmetric 
$e^+ e^-$ collider. The $CP$ asymmetries obtained for 
$D_{CP}K$ are: ${\cal{A}}_1 = 0.06 \pm 0.14 (\rm stat) \pm 0.05 (\rm sys)$,
${\cal{A}}_2 = -0.12 \pm 0.14 (\rm stat) \pm 0.05 (\rm sys)$
and for $D^*_{CP}K$: ${\cal{A}}_1^* = -0.20 \pm 0.22 (\rm stat) 
\pm 0.04 (\rm sys)$,
${\cal{A}}_2^* = 0.13 \pm 0.30 (\rm stat) \pm 0.08 (\rm sys)$.

\end{abstract}

\maketitle 
{\renewcommand{\thefootnote}{\fnsymbol{footnote}}}
\setcounter{footnote}{0}
\normalsize

\par
\begin{figure}[htbp]
\begin{center}
\includegraphics[width=0.8\textwidth]{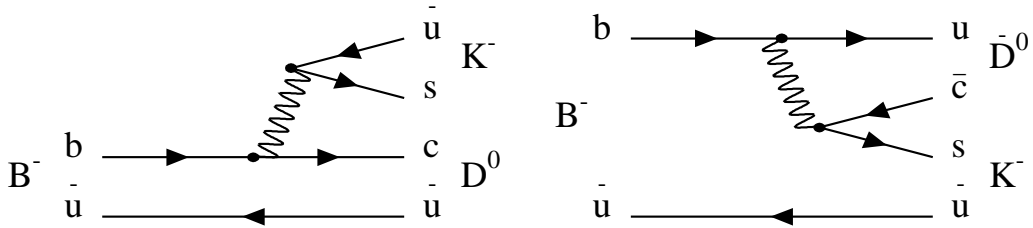}
\end{center}
\caption{Feynman diagrams for $B^{-} \rightarrow D^{0}K^{-}$ and 
$B^{-}\rightarrow\bar{D}^{0}K^{-}$.}
\label{feynman}
\end{figure}

 Measurements of the decay rates of $B^\pm \to D^{(*)}K^\pm$
provide a theoretically clean method for extracting the Unitarity
Triangle angle $\phi_3$, an angle in the Cabibbo-Kobayashi-Maskawa (CKM)
quark mixing matrix~\cite{ckm}. Since both a $D^0$ and a $\bar{D}^{0}$
can decay into the same $CP$ eigenstate ($D_{CP}$, or $D_1$ for a 
$CP$-even state and $D_2$ for a $CP$-odd state),
the $b\rightarrow c$ and $b\rightarrow u$ processes shown in 
Fig.~\ref{feynman} interfere in the 
$B^{\pm} \to D_{CP} K^{\pm}$ decay channel. This interference may lead to
direct $CP$ violation. To measure $D$ meson decays to $CP$ eigenstates
a large number of $B$ meson decays is required since the branching fractions
to these modes are of order 1\%. To extract $\phi_3$ using the 
GLW method~\cite{gwl}, the following 
observables sensitive to $CP$ violation must be measured: the
asymmetries 
\begin{eqnarray}
\label{eq1}
\small
{\cal{A}}_{1,2} & \equiv & \frac{{\cal B}(B^- \rightarrow D_{1,2}K^-) -
{\cal B}(B^+ \rightarrow D_{1,2}K^+) }{{\cal B}(B^- \rightarrow
D_{1,2}K^-) + {\cal B}(B^+ \rightarrow D_{1,2}K^+) }\\ 
& = & \frac{2 r \sin \delta ' \sin \phi_3}{1 + r^2 + 2 r \cos \delta '
\cos \phi_3}
\end{eqnarray}
and the double ratios 
\begin{eqnarray}
\label{eq2}
{\cal{R}}_{1,2} & \equiv & \frac{R^{D_{1,2}}}{R^{D^{0}}}  = 1 + r^2 + 2 r
\cos \delta ' \cos \phi_3\\
\delta ' & = & \left\{
             \begin{array}{ll}
              \delta & \mbox {{\rm  for }$D_1$}\\
              \delta + \pi&  \mbox{{\rm for }$D_2$}\\
             \end{array}
             \right.
\end{eqnarray}

\noindent The ratios $R^{D_{1,2}}$ and $R^{D^{0}}$ are defined as

\begin{eqnarray*}
\small
R^{D_{1,2}} & = & \frac{{\cal B}(B^- \rightarrow D_{1,2}K^-)+{\cal B}(B^+
\rightarrow D_{1,2}K^+)}{{\cal B}(B^- \rightarrow D_{1,2}\pi^-) +
{\cal B}(B^+ \rightarrow D_{1,2}\pi^+)}\\
	R^{D^{0}} & = & \frac{{\cal
B}(B^- \rightarrow D^{0} K^-)+{\cal B}(B^+ \rightarrow
\bar{D}^{0} K^+)}{{\cal B}(B^- \rightarrow D^{0}\pi^-) + {\cal B}(B^+
\rightarrow \bar{D}^{0} \pi^+)}
\end{eqnarray*}

\noindent where $r \equiv |A(B^- \to \bar{D}^0 K^-)/A(B^- \to D^0 K^-)|$ 
is the ratio of the magnitudes of the two tree diagrams
shown in Fig.~\ref{feynman}, $\delta$ is their strong-phase
difference. 
The ratio $r$ is given by the product of CKM factors and a color 
suppression factor, that characterizes the magnitude of $CP$
asymmetry. The asymmetries
and double ratios can be calculated for $D^*$ in a similar manner 
(notation ${\cal{A}}^*_{1,2}$ and ${\cal{R}}^*_{1,2}$). Here we have
assumed that mixing and $CP$ violation in the neutral $D$ meson
system can be neglected.
%$D^0-\bar{D}^0$ mixing can be neglected.

\par Previously, Belle~\cite{sugi} and BaBar~\cite{babar1} reported 
the observation of the decays $B^{-}\rightarrow D_{1}K^{-}$ and 
$B^{-} \rightarrow D_{2}K^{-}$. BaBar~\cite{babar2} also reported 
the observation of the decay $B^{-}\rightarrow D_{1}^{*}K^{-}$. This
paper reports more precise measurements of the 
$B^{-}\rightarrow D_{CP}K^{-}$ channels, superseding our previous 
result, and a study of the 
decays $B^- \rightarrow D^{*}_1 K^-$ and 
$B^- \rightarrow D^{*}_2 K^-$ with a data sample
corresponding to 275 $\times 10^6$ $B\overline{B}$ pairs.

%%%%%%%%%%%%%%%%%%%%%%%%%%%%%%%%%%%%%%%%%%%%%%%%%%%%%%%%%%%%%%%%%%

\par The Belle detector is a large-solid-angle magnetic
spectrometer that consists of a silicon vertex detector (SVD),
a 50-layer central drift chamber (CDC), an array of
aerogel threshold \v{C}erenkov counters (ACC),
a barrel-like arrangement of time-of-flight
scintillation counters (TOF), and an electromagnetic calorimeter (ECL)
comprised of CsI(Tl) crystals located inside
a superconducting solenoid coil that provides a 1.5~T
magnetic field.  An iron flux-return located outside of
the coil is instrumented to detect $K_L^0$ mesons and to identify
muons (KLM).  The detector is described in detail elsewhere~\cite{Belle}.
%Two different inner detector configurations were used. For the first sample
%of 152 $\times 10^6$ $B\bar{B}$ pairs, a 2.0 cm radius beampipe
%and a 3-layer silicon vertex detector were used;
%for the latter 123 $\times 10^6$ $B\bar{B}$ pairs,
%a 1.5 cm radius beampipe, a 4-layer silicon detector
%and a small-cell inner drift chamber were used\cite{Ushiroda}.
%%%%%%%%%%%%%%%%%%%%%%%%%%%%%%%%%%%%%%%%%%%%%%%%%%%%%%%%%%%%%%%%%%

%\section{Event Reconstruction}
%%%%%%%%%%%%%%%%%%%%%%%%%%%%%%%%%%%%%%%%%%%%%%%%%%%%%%%%%%%%%%%%%%

\par $D^{0}$ mesons are reconstructed in the Cabibbo-favored modes 
($D_{f}$)~\cite{conjugate}: 
$K^{-}\pi^{+}$, $CP$-even modes ($D_{1}$): $K^{+}K^{-}$, $\pi^{-}\pi^{+}$ 
and $CP$-odd modes ($D_{2}$): $K_{S}^0 \pi^{0}$, $K_{S}^0 \omega$, $K_{S}^0 \phi$. 

\par Neutral pions are reconstructed from pairs of photons, 
selected in the invariant mass range $118 \ {\rm MeV}/c^2 < 
M(\gamma\gamma) < 150 \ {\rm MeV}/c^2$, corresponding
to a $\pm 2.5 \sigma$ window, where $\sigma$ is 
the $\pi^0$ mass resolution. Each photon is required to have energy greater
than 30 MeV in the laboratory frame, and also in this frame the 
pion candidate's momentum must exceed 0.5 (0.1) GeV/$c$ for the
$K_{S}^0 \pi^{0}$ ($K_{S}^0 \omega$) mode.
The $\pi^{0}$ candidates are kinematically 
constrained to the nominal $\pi^{0}$ mass.
%In addition, the $\chi^{2}$ obtained from the
%kinematical fit of the $\pi^{0}$ is required to be smaller than 25. 

\par Each charged track not coming from a $K_{S}^0$ candidate is required 
to be consistent with coming from the interaction point (IP). 
For each charged track, information from the ACC, TOF and CDC is used
to identify the particle (PID) as either a pion or kaon 
via the $K/\pi$ likelihood ratio $P(K/\pi)= L_K/(L_K+L_\pi)$,
where $L_K$ and $L_\pi$ are kaon and pion likelihoods. 
With the exception of the prompt kaon from the $B$ meson decay
(``fast track''), all kaon candidates must satisfy the 
PID requirement, $P(K/\pi) > 0.3$.
This requirement selects kaons with an efficiency
of 92\% and a pion misidentification rate of 18\%.
\par The $K_{S}^0$ candidates are formed from two oppositely charged pions
with an invariant mass 
required to be within $8.5 \ {\rm MeV}/c^2$ of the nominal mass
($\sim 3 \sigma$). The $\phi$ meson is
reconstructed from two oppositely charged kaons in a mass window
$|M(K^{+}K^{-})-m_{\phi}| < 10$ MeV/c$^{2}$. $\omega$ mesons are
reconstructed from $\pi^{+}\pi^{-}\pi^{0}$ combinations in the mass
window 0.757 GeV/$c^2 < M(\pi^{+}\pi^{-}\pi^{0}) <$ 0.82 GeV/$c^2$; 
a loose requirement of $P(K/\pi) < 0.9$ is applied to 
the charged pions in the $\omega$ candidate. 

\par For the $D$ candidates,
a $3\sigma$ mass requirement is applied, where $\sigma$ is the $D$ mass 
resolution which ranges from 5 to 12 MeV. 
$D^{*}$ candidates are reconstructed in the $D\pi^0$ 
decay channel depending on the decay channel. 
The mass difference between $D^{*}$ and $D$ candidates
is required to be within 2.8 MeV/$c^2$ ($\sim 3 \sigma$) of the nominal
value~\cite{pdg}.
$B$ meson candidates are formed by combining the $D^{(*)}$ candidates with
one charged track (denoted $h^\pm$). 
The signal is identified by two kinematic variables: 
the beam-constrained mass $M_{\rm bc}$ and the
energy difference $\Delta E$ calculated in the $\Upsilon (4S)$ 
center of mass (CM) frame, 
$M_{\rm bc} \equiv \sqrt{E_{\rm beam}^{2} - 
|\vec{p}_{D}+\vec{p}_{h}|^{2}}$ and
$\Delta E \equiv E_{D} + E_{h} -E_{\rm beam}$, where $E_{\rm beam}$ is 
the beam energy, $\vec{p}_{D}$ and $E_{D}$ are the momentum and 
energy of the $D^0$ candidate and $\vec{p}_{h}$ and $E_{h}$ are 
the momentum and energy of the $K^-/\pi^-$ candidate assuming
the pion mass. With this definition, $B^- \to D^0 \pi^-$ events
peak at $\Delta E = 0$, while $B^- \to D^0 K^-$ events peak around
$\Delta E = -49$ MeV.
Signal candidates are selected with $M_{\rm bc} > 5.2$ GeV/$c^2$ 
and $|\Delta E| < 0.2$ GeV. The
experimental resolution for $M_{\rm bc}$ is approximately $3 \ {\rm MeV}$, 
dominated by the beam energy spread.
The $\Delta E$ resolution is typically $10 \ {\rm MeV}$ 
for all-charged-particle final states ($D_{1}$ modes). For final
states with photons or neutral pions, the $\Delta E$ 
resolution becomes broader and somewhat skewed to negative values.

%%%%%%%%%%%%%%%%%%%%%%%%%%%%%%%%%%%%%%%%%%%%%%%%%%%%%%%%%%%%%%%%%%

\par Event topology is used to distinguish $B\bar{B}$ events from
continuum background.  At the $\Upsilon(4S)$, the two $B$ mesons
are produced nearly at rest so these events tend to be spherical, whereas
continuum events have a two-jet topology.  
We construct a Fisher discriminant~\cite{Fisher} of modified 
Fox-Wolfram moments
called the Super-Fox-Wolfram ($SFW$)~\cite{SFW}, 
where the Fisher coefficients are optimized by 
maximizing the separation between $B\bar{B}$ events and continuum
events.
The angle in the CM frame between the $B$ flight direction and the 
beam axis, $\cos \theta_{B}$, is also used.
These two  independent variables ($SFW$ and $\cos \theta_{B}$) are
combined to form a likelihood ratio: ${\cal R} =
L_{\rm sig}/(L_{\rm sig}+L_{\rm cont})$, where $L_{\rm sig}$ 
and $L_{\rm cont}$ are
defined as the product of $SFW$ and $\cos \theta_{B}$ likelihood. 
The ${\cal R}$ requirement is optimized for 
each submode of $DK$ and $D^*K$ using $N_S/\sqrt{N_S+N_B}$, where 
$N_S$($N_B$) is the expected number of signal (background) events 
in the signal region (the coefficients used for the Fisher 
discriminant are common to all the sub-modes). 
The expected number of signal events is obtained assuming the
branching ratio given in the Review of Particle Physics~\cite{pdg}.
Since the continuum background is negligible for the $K_{S}^0 \phi$ mode,
we do not apply an ${\cal R}$ requirement.
\par For events with more than one candidate (1\%--2\% for all modes, 
except for $K_S^0 \omega$, $\sim$10\%), a single
candidate is selected on the basis of a $\chi^{2}$ determined from
the difference between the measured and nominal values of masses 
($D$, $D^*$, $K^0_S$, $\omega$, $\phi$ masses) and then 
the highest ${\cal R}$ value. 

%%%%%%%%%%%%%%%%%%%%%%%%%%%%%%%%%%%%%%%%%%%%%%%%%%%%%%%%%%%%%%%%%%

\par Signal yields are obtained from fitting the $\Delta E$ distributions for 
$5.27 \ {\rm GeV}/c^2 < M_{\rm bc} < 5.29 \ {\rm GeV}/c^2$. The PID
for the fast $\pi$ or $K$ is used to distinguish 
between $D^{(*)}\pi$ and $D^{(*)}K$ modes (with a requirement $P(K/\pi) > 0.8$ for 
$D^{(*)}K$, which selects kaons with an efficiency
of 80\% and a pion misidentification rate of 7\%,
and the remainder as $D^{(*)}\pi$). Signal peaks are fitted with 
double Gaussians.
% where the shape is taken from $D\pi$ in data and used in $DK$. 
Shifts in the mean position and differences in resolution
seen between Monte Carlo (MC) and data in $D\pi$ are used to 
correct the $DK$ fits in data.
The continuum background is modeled by a first order polynomial whose slope 
is obtained from the $M_{\rm bc}$ sideband ($M_{\rm bc} <$ 5.25 GeV/$c^2$). 
Backgrounds from $B$ meson decays are modeled by large MC samples 
using a smoothed histogram. 
When statistics are small, as is the case in $D^*K$, shapes from
$D^*\pi$ are used directly.

\par Backgrounds are studied using MC samples for known backgrounds and $D^0$
sidebands in data. A peaking background is found
for $B \rightarrow D K$ (where $D \rightarrow \pi\pi$) coming from 
$B \rightarrow D \pi$  ($D \rightarrow K\pi$), which is suppressed by making
a $3 \sigma$ mass requirement on the $K\pi$ invariant mass.
\par For $DK$, in the $K^+K^-$ and $\pi^+\pi^-$ modes,
clear peaks are seen in the $D$ mass sideband, defined as 
1.80 GeV/$c^2 < M(hh) < 1.83$ GeV/$c^2$ and 
1.90 GeV/$c^2 < M(hh) < 1.93$ GeV/$c^2$, 
where $h$ is a charged kaon or pion. 
These peaks come from the $B \rightarrow KKK$ and 
$B \rightarrow K\pi\pi$ modes,
respectively. The yields obtained ($63.5\pm7.5$ events) are in agreement 
with the results of a dedicated study of these channels~\cite{garmash} 
and allow an estimate of the peaking backgrounds for these modes.  
The sidebands are scaled (factor 0.5) and subtracted from the yields 
of $D_{1}$, and hence are taken into account in the asymmetries
and double ratios defined in Eqns.~\ref{eq1}-\ref{eq2}. Note that such 
effects are not seen in $B \rightarrow D^*K$ since the $D^*$ provides 
an extra constraint to reduce these backgrounds.

\par Backgrounds in the $D_2$ modes, $K_S^0\omega$ and $K_S^0\phi$,
need careful consideration because they can be modes of non-$CP$ or with 
opposite $CP$ (opposite asymmetry) to the mode considered. 
Possible backgrounds to $K_S^0\omega$ include non-$CP$ modes 
$K_S^0\pi\pi\pi^0$ and $K^*\rho$, and backgrounds to $K_S^0\phi$ 
include non-$CP$ modes $a_0^\pm(980)K^\mp$, $K_S^0KK$, 
and opposite $CP$ modes $K_S^0a_0^0(980)$ and $K_S^0f_0(980)$. 
To determine contributions from these backgrounds, 
the data $\Delta E$ distributions for $D\pi$ modes are fitted
in bins of $\omega$ or $\phi$ helicity angle.
The helicity angle 
$\theta_{\rm hel}$ for $\phi$ ($\omega$) is defined
as the angle between one of the kaons from $\phi$ 
(the normal to the $\omega$ decay plane) and $D$ momentum 
in the $\phi$ ($\omega$) rest frame.  
The yields as a function of helicity angle are 
then fitted allowing for possible 
contributions from signal and either $K_S^0\pi\pi\pi^0$ and $K_S^0KK$. 
The fraction of signal is estimated to be $88.8 \pm 8.4\%$
for $K_S^0\omega$ and $84.0 \pm 12.5\%$ for $K_S^0\phi$. When a helicity
requirement ($|\cos \theta_{\rm hel}| > 0.4$) is imposed to further 
reduce the backgrounds, 
these fractions become $92.4 \pm 9.8\%$ for $K_S^0\omega$
and $88.6 \pm 11.1 \%$ for $K_S^0\phi$. 

\par
\begin{figure}[htbp]
\begin{center}
\begin{tabular}{cc}
\includegraphics[width=0.4\textwidth]{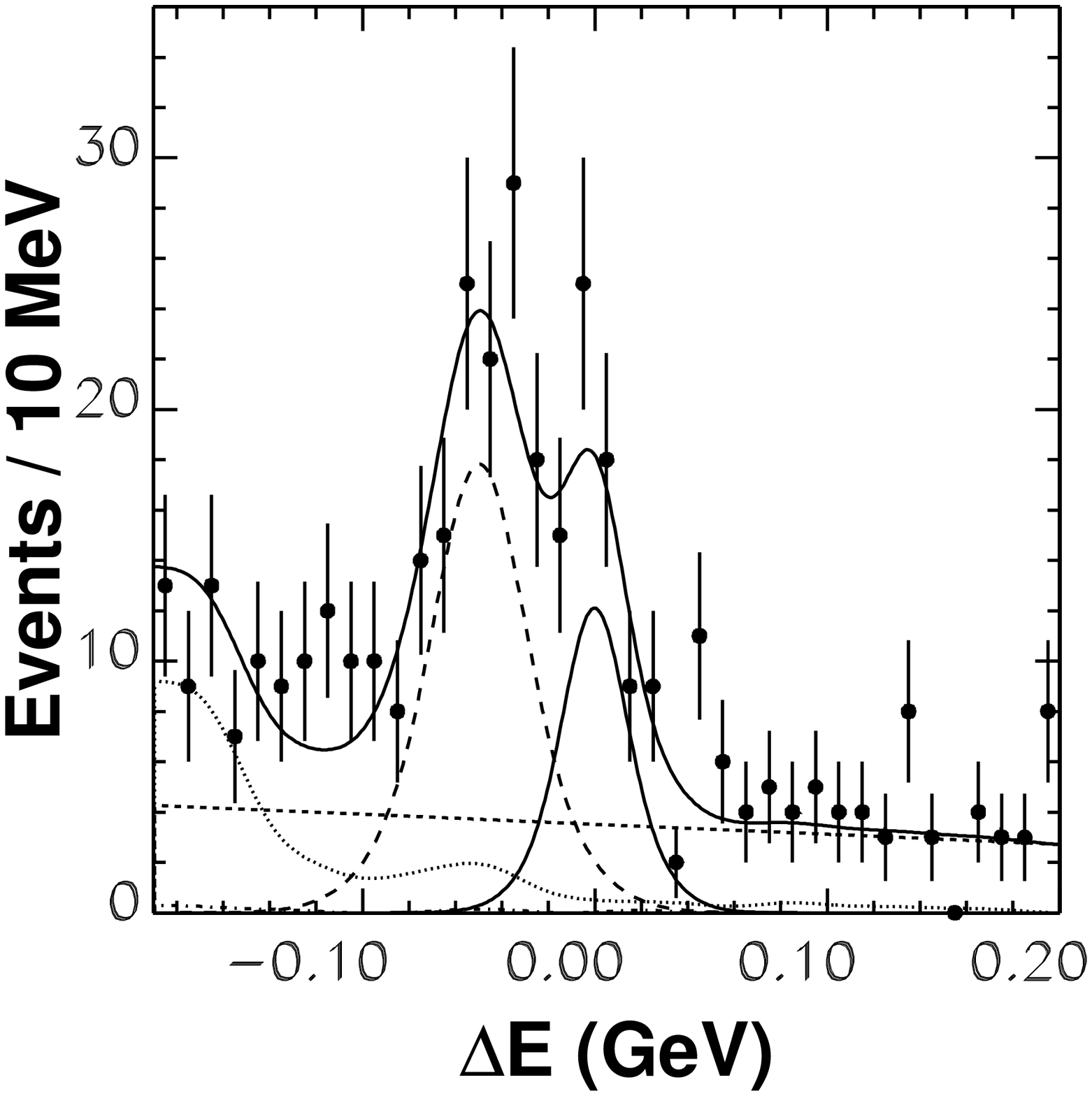} &
\includegraphics[width=0.4\textwidth]{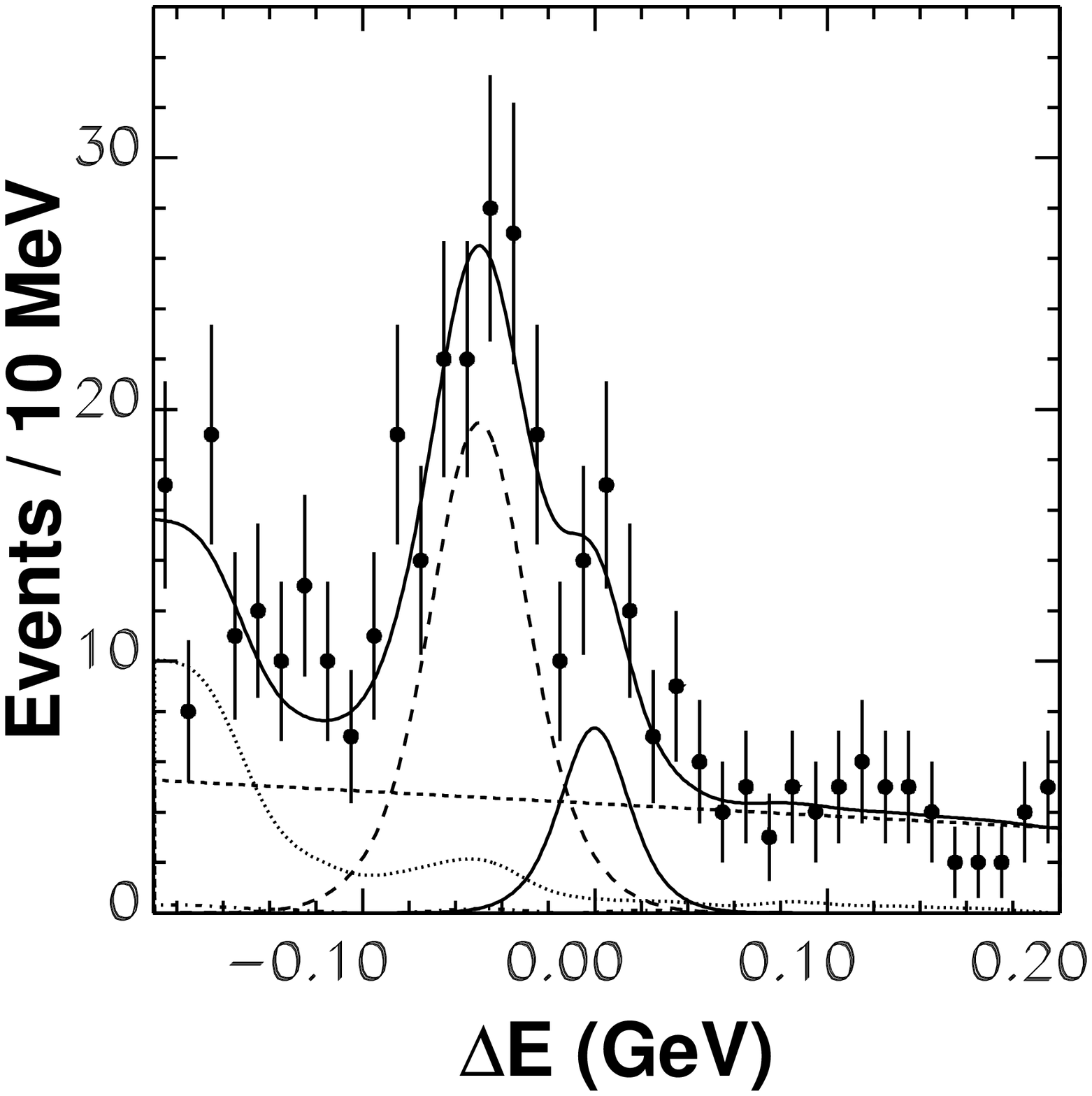} \\
\includegraphics[width=0.4\textwidth]{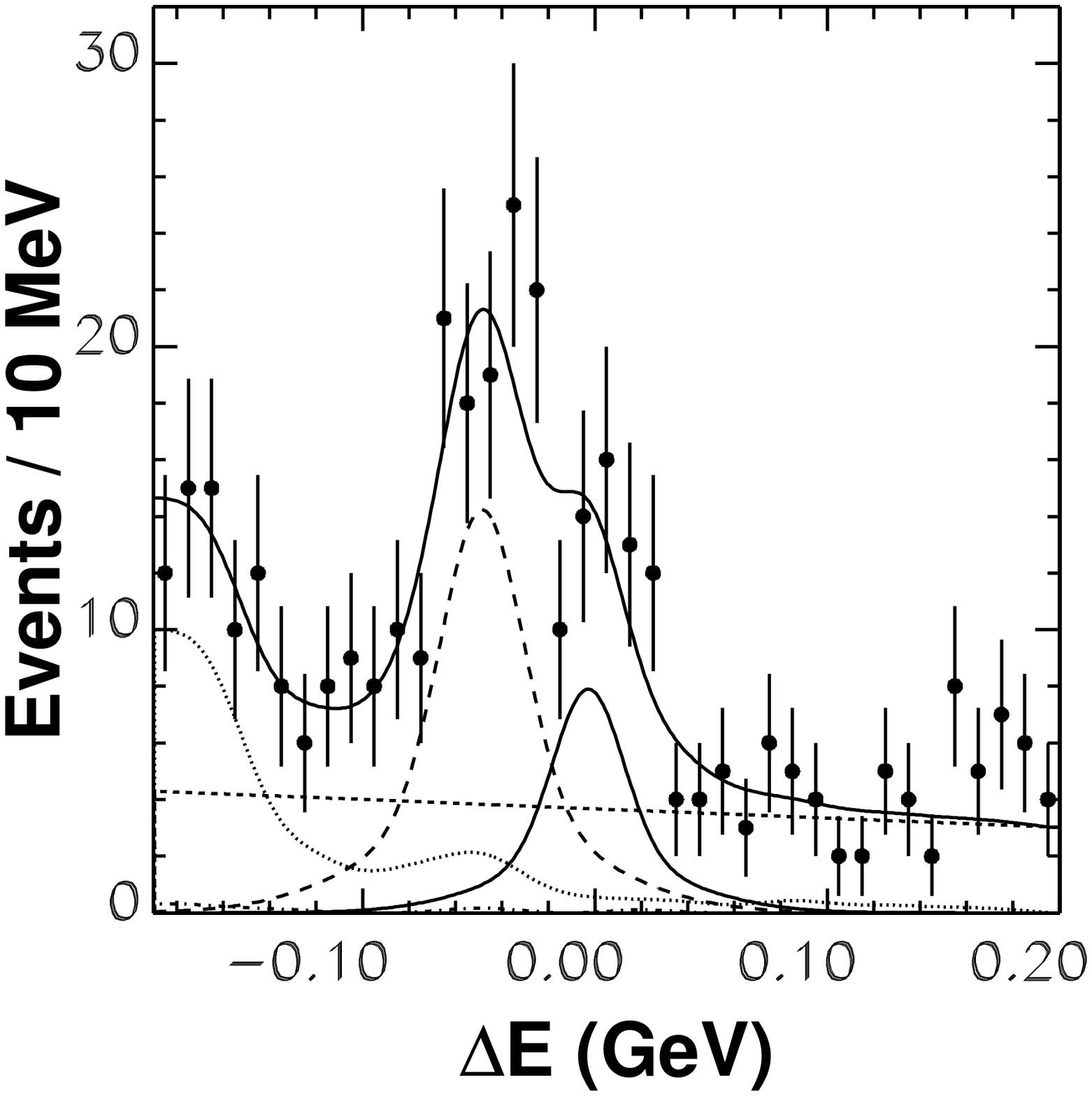} &
\includegraphics[width=0.4\textwidth]{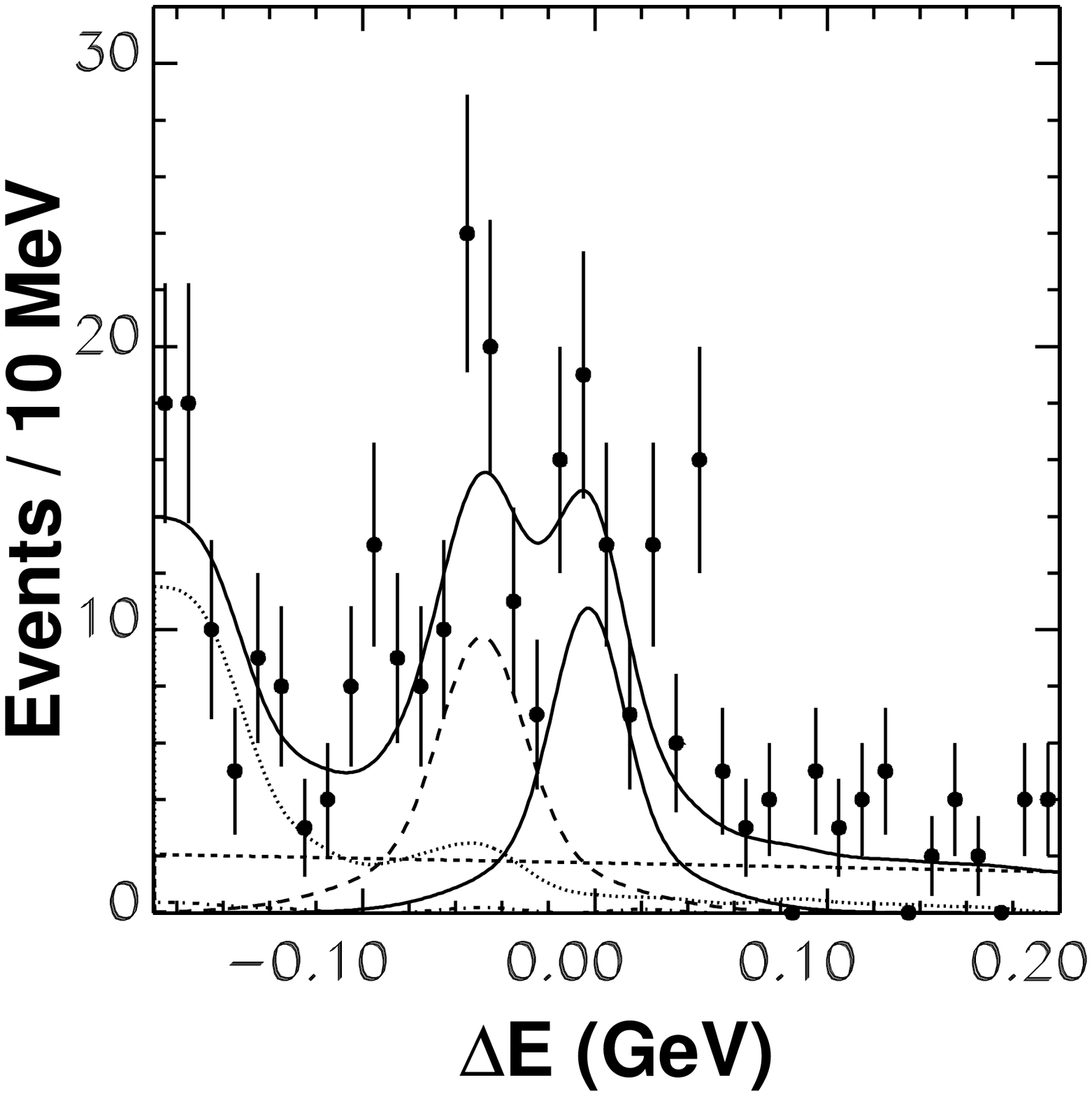} \\
\end{tabular}
\end{center}
\caption{$\Delta E$ distributions for (top left) $B^+ \rightarrow D_1 K^+$, 
(top right) $B^-\rightarrow D_1 K^-$, (bottom left) 
$B^+ \rightarrow D_2 K^+$, (bottom right) 
$B^-\rightarrow D_2 K^-$. Points with error bars are the data 
and the solid lines show the fit results.
The components of the fit are the background from $B$ meson decays 
(dotted line), the continuum background (dashed), 
the signal $DK$ (left) and $D\pi$ (right).}
\label{dk2}
\end{figure}

 The fitted $\Delta E$ distributions for positively and negatively 
charged $B$ meson candidates are shown in Fig.~\ref{dk2}. 
Table~\ref{table:asym_dh} gives the corresponding 
yields and asymmetries with their statistical uncertainties. 
The asymmetries in the control samples ($D_{f}$)
are consistent with zero, as expected. 
The modes of interest are $D_1K$ and $D_2K$: the $B^+$ and $B^-$ 
yields are used to calculate asymmetries after
peaking background subtraction for $D_1$. 
For $D_2$ modes, the asymmetry is estimated mode by mode and the dilution 
factor due to the $CP$ content of the background is taken into 
account, assuming no $CP$ for $K_S^0 \omega$ 
background and opposite $CP$ for $K_S^0\phi$ background.

\begin{table*}[htbp]
\caption{Yields and asymmetries obtained for $Dh$ and $D^*h$ modes.
For $D_2$ modes, the asymmetry is estimated mode by mode taking into
account the $CP$ content of the background.}
\begin{ruledtabular}
\begin{tabular}{lcccc}
 &  $\sum B$  & $B^{+}$  & $B^{-}$ & $\cal A$\\
\hline
$D_f\pi$ & 19266$\pm$150 &      9677$\pm$103 & 9521$\pm$102 & $-$0.008$\pm$0.008\\
$D_1\pi$ & 2163$\pm$56 & 1049$\pm$38 & 1124$\pm$37 & 0.035$\pm$0.024\\
$D_2\pi$ & 2168$\pm$61 & & & 0.017$\pm$0.026\\
$D_fK$ & 1131$\pm$41 & 528$\pm$28 & 603$\pm$29 & 0.066$\pm$0.036\\
\hline
$D_1K$ & 143.3$\pm$21.9 & 70.2$\pm$14.7 & 79.2$\pm$15.7 & 0.060$\pm$0.144$\pm$0.046\\
$D_2K$ &  149.5$\pm$19.0 & & & $-$0.117$\pm$0.141$\pm$0.049\\
\hline
\hline
$D^*\pi$ & 5434$\pm$101 & 2756$\pm$59 & 2678$\pm$59 & $-$0.014$\pm$0.015\\
$D^*_{1}\pi$ & 662$\pm$37 & 338$\pm$21 & 322$\pm$21 & $-$0.021$\pm$0.045\\
$D^*_{2}\pi$ & 604$\pm$38 &    &                       & $-$0.090$\pm$0.051\\
$D^* K$ & 256$\pm$22 & 140$\pm$16 & 117$\pm$15 & $-$0.089$\pm$0.086       \\
\hline
$D^*_{1} K$ & 43.9$\pm$10.2 & 27.3$\pm$7.4 & 18.2$\pm$6.9 & $-$0.200$\pm$0.224$\pm$0.035\\
$D^*_{2} K$ & 32.7$\pm$10.0 & & & 0.131$\pm$0.300$\pm$0.076\\
\end{tabular}
\end{ruledtabular}
\label{table:asym_dh}
\end{table*}

\par
\begin{figure}[htb]
\begin{center}
\begin{tabular}{cc}
\includegraphics[width=0.4\textwidth]{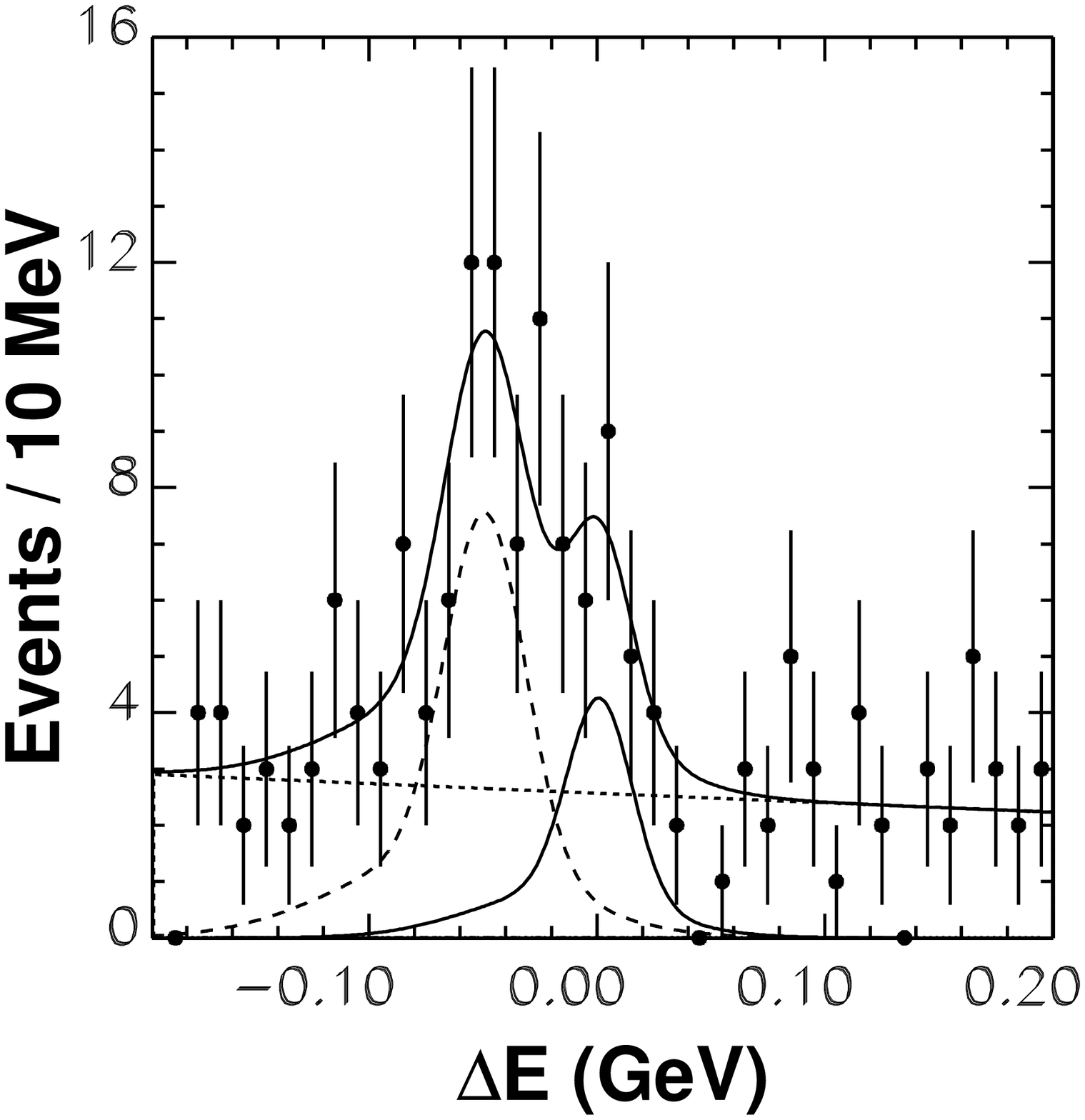}
\includegraphics[width=0.4\textwidth]{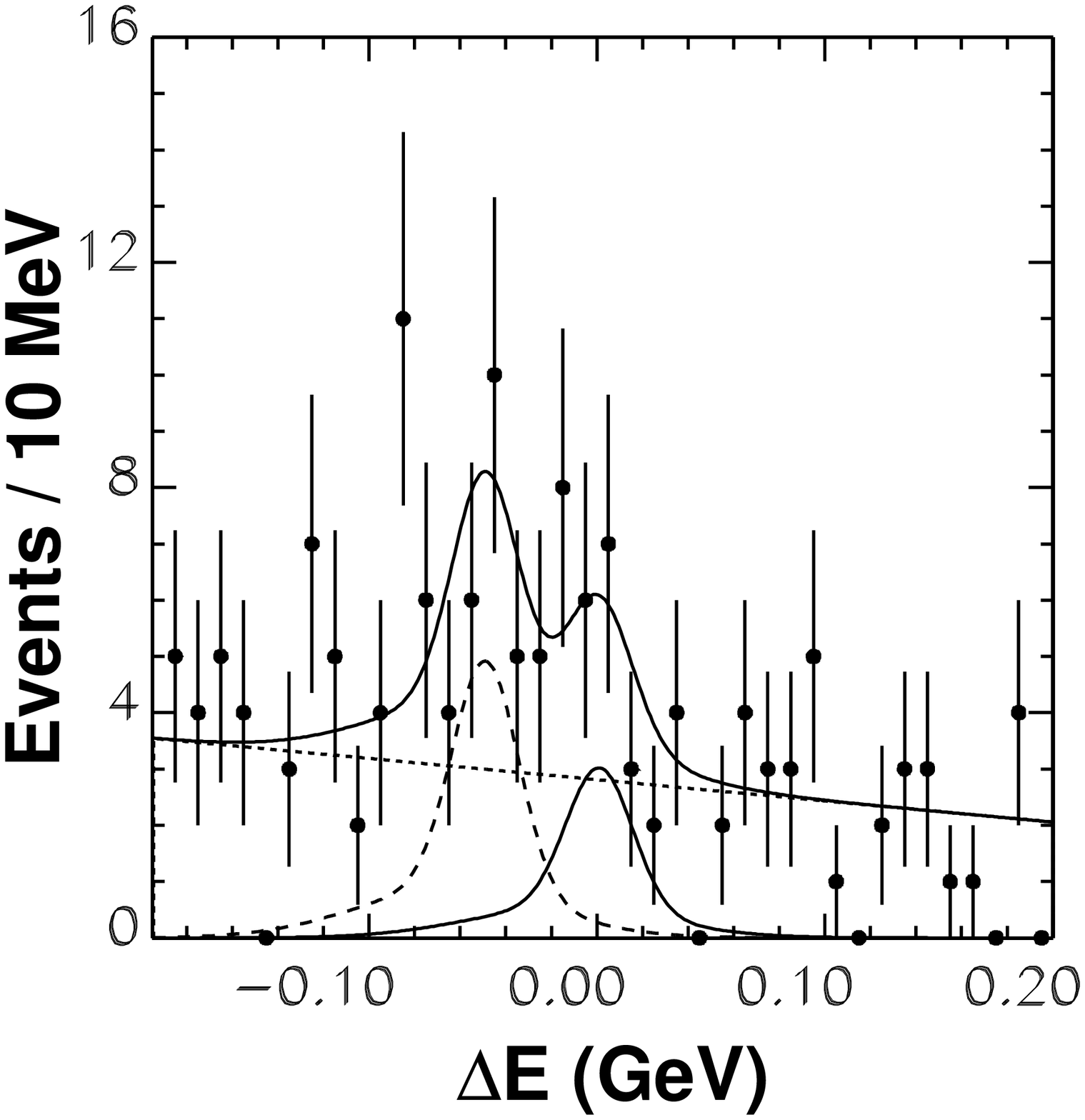}
\end{tabular}
\end{center}
\caption{$\Delta E$ distributions for (left) $B^\pm \rightarrow
D^{*0}_1K^\pm$ and (right) $B^\pm \rightarrow
D^{*0}_2K^\pm$. Points with error bars are the data 
and the solid lines show the fit results.}
\label{dstk}
\end{figure}

 The yields obtained for $D^*_{1} K$ and  $D^*_{2} K$ (Fig.~\ref{dstk}) 
are $43.9 \pm 10.2$ and $32.7 \pm 10.0$ respectively, which correspond to 
significances of 5.2$\sigma$ and 3.3$\sigma$ ($K_S^0\pi^0$ 2.9$\sigma$,
$K_S^0\omega$ 0.9$\sigma$, $K_S^0\phi$ 1.4$\sigma$) where the 
significance is calculated as $\sqrt{-2 \ln
({\cal L}_{0}/{\cal L}_{\rm max})}$, where ${\cal L}_{\rm max}$ and ${\cal
L}_{0}$ denote the maximum likelihood with the nominal signal yield
and with signal yield fixed to 0, respectively.

\par The sources of systematic errors for the double ratios come from
the uncertainty in yield extraction, uncertainty in signal fractions for 
$K_S^0\omega$ and $K_S^0\phi$ (1\%) and the
uncertainty in the contributions of peaking background
from $D$ sideband data. The uncertainty in yield extraction is
estimated by varying the fitting parameters, such as the slope used
for continuum or widths and means for Gaussians used for signals by
$\pm 1 \sigma$ (6\%--8\%). The uncertainty due to peaking $D$ sidebands is
taken from the error on the estimated contribution: 6\% for
${\cal{R}}_1$. 
These errors are added in quadrature for $D^{(*)}_{1}$
and $D^{(*)}_{2}$.

\par Systematic errors for ${\cal A}$ are from intrinsic detector
charge asymmetry, measured from
the control samples $B\rightarrow D_{f}\pi$, 
(0.02), uncertainty in signal fraction for $K_S^0\omega$ 
and $K_S^0\phi$ and on the $CP$ content assumption of the
peaking background (0.01), yield extraction (0.02--0.04) and PID (0.01).

\par The asymmetries for $D_{1,2}K$, ${\cal A}_{1}$ and ${\cal A}_{2}$, are
found to be:

\begin{center}
\begin{tabular}{lcc}
${\cal A}_1$ & = &  0.06 $\pm$ 0.14(stat) $\pm$ 0.05(sys)\\
${\cal A}_2$ & = & $-$0.12 $\pm$ 0.14(stat) $\pm$ 0.05(sys).
\end{tabular}
\end{center}

The double ratios are:

\begin{center}
\begin{tabular}{lcc}
${\cal{R}}_1$ & = & 1.13 $\pm$ 0.16(stat) $\pm$ 0.08(sys) \\
${\cal{R}}_2$ & = & 1.17 $\pm$ 0.14(stat) $\pm$ 0.14(sys). \\
\end{tabular}
\end{center}

\par The asymmetries for $D^*_{1,2}K$ are found to be:
\begin{center}
\begin{tabular}{lcc}
${\cal A}^*_1$ & = & $-$0.20 $\pm$ 0.22(stat) $\pm$ 0.04(sys) \\
${\cal A}^*_2$ & = & 0.13 $\pm$ 0.30(stat) $\pm$ 0.08(sys), \\
\end{tabular}
\end{center}

\noindent where the systematic errors are calculated 
in a similar way to the $Dh$ case. The double ratios found are:

\begin{center}
\begin{tabular}{lcc}
${\cal{R}}^*_1$ & = & 1.41 $\pm$ 0.25(stat) $\pm$ 0.06(sys) \\
${\cal{R}}^*_2$ & = & 1.15 $\pm$ 0.31(stat) $\pm$ 0.12(sys).
\end{tabular}
\end{center}

%%%%%%%%%%%%%%%%%%%%%%%%%%%%%%%%%%%%%%%%%%%%%%%%%%%%%%%%%%%%%%%%%%

%\section{Discussion}

\par For each state such as $B \rightarrow D_{CP}K$ four
observables are measured (${\cal A}_1$, ${\cal A}_2$, ${\cal R}_1$, ${\cal
R}_2$). Since these are related by ${\cal A}_1 {\cal R}_1 = -{\cal A}_2
{\cal R}_2$ there are then only three independent observables and there 
are also
three physics quantities that should be extracted, $\phi_3$, $r_{DK}$ and
$\delta_{DK}$. The measured asymmetries are consistent within errors with
zero and also with the Standard Model expectation. These measurements,
while not sufficiently accurate to provide a measurement of $\phi_3$,
can be used to constrain $\phi_3$ through a global fit~\cite{ckmfitter}, 
and the addition of modes such as $D^*_{CP} K$ can add further 
constraints~\cite{soni}.

%%%%%%%%%%%%%%%%%%%%%%%%%%%%%%%%%%%%%%%%%%%%%%%%%%%%%%%%%%%%%%%%%%

\par In summary, using 275 $\times 10^6$ $B\bar{B}$ events this paper 
reports results from the decays $B^\pm \rightarrow D_{CP}K^\pm$ and
$B^\pm \rightarrow D^*_{CP}K^\pm$ where the $CP$ eigenstates are those
of the $D$ meson. The partial rate asymmetries ${\cal A}_{1,2}$ are
measured in both cases and are consistent with zero. The study of
$B \rightarrow D_{CP} K$ is made with three times the statistics of 
previous studies.
$B \rightarrow D^*_1 K$ and $B \rightarrow D^*_2 K$ are observed  
and their asymmetries and double ratios are also measured.

%%%%%%%%%%%%%%%%%%%%%%%%%%%%%%%%%%%%%%%%%%%%%%%%%%%%%%%%%%%%%%%%%%

\section*{Acknowledgments}
%***** Acknowledgments *****
% Please paste this acknowledgement into your latex file. 
 %----------- Long version, for most papers ----------- 
%We thank the KEKB group for the excellent operation of the
%accelerator, the KEK Cryogenics group for the efficient
%operation of the solenoid, and the KEK computer group and
%the National Institute of Informatics for valuable computing
%and Super-SINET network support. We acknowledge support from
%the Ministry of Education, Culture, Sports, Science, and
%Technology of Japan and the Japan Society for the Promotion
%of Science; the Australian Research Council and the
%Australian Department of Education, Science and Training;
%the National Science Foundation of China under contract
%No.~10175071; the Department of Science and Technology of
%India; the BK21 program of the Ministry of Education of
%Korea and the CHEP SRC program of the Korea Science and
%Engineering Foundation; the Polish State Committee for
%Scientific Research under contract No.~2P03B 01324; the
%Ministry of Science and Technology of the Russian
%Federation; the Ministry of Education, Science and Sport of
%the Republic of Slovenia; the National Science Council and
%the Ministry of Education of Taiwan; and the U.S.\
%Department of Energy.

%-------- Short version, if necessary, for PRL -----------
% currently commented out
We thank the KEKB group for the excellent operation of the
accelerator, the KEK cryogenics group for the efficient
operation of the solenoid, and the KEK computer group and
the NII for valuable computing and Super-SINET network
support.  We acknowledge support from MEXT and JSPS (Japan);
ARC and DEST (Australia); NSFC (contract No.~10175071,
China); DST (India); the BK21 program of MOEHRD and the CHEP
SRC program of KOSEF (Korea); KBN (contract No.~2P03B 01324,
Poland); MIST (Russia); MESS (Slovenia); NSC and MOE
(Taiwan); and DOE (USA).


\begin{thebibliography}{99}
\bibitem{conjugate}
Hereafter, the inclusion of the charge conjugate mode decay
is implied unless otherwise stated.
\end{thebibliography}
\end{document}